\documentclass[aps, prb, twocolumn, showpacs]{revtex4}
\usepackage{amsmath,amsfonts, amssymb, amsthm}
\usepackage{bm}
\usepackage{mathrsfs}
\usepackage{graphicx, subfigure}
\usepackage{subfigure}
\usepackage{verbatim}
\usepackage{mathrsfs}
\usepackage{hyperref}

\begin{document}

\newcommand{\odiff}[2]{\frac{\di #1}{\di #2}}
\newcommand{\pdiff}[2]{\frac{\partial #1}{\partial #2}}
\newcommand{\di}{\mathrm{d}}
\newcommand{\ii}{\mathrm{i}}
\newcommand{\ua}{\uparrow}
\newcommand{\da}{\downarrow}
\renewcommand{\vec}[1]{{\mathbf #1}}
\newcommand{\vx}{{\bm x}}
\newcommand{\ket}[1]{|#1\rangle}
\newcommand{\bra}[1]{\langle#1|}
\newcommand{\pd}[2]{\langle#1|#2\rangle}
\newcommand{\tpd}[3]{\langle#1|#2|#3\rangle}
\renewcommand{\vr}{{\vec{r}}}
\newcommand{\vk}{{\vec{k}}}
\renewcommand{\ol}[1]{\overline{#1}}
\newtheorem{theorem}{Theorem}
\newcommand{\comments}[1]{}
\newcommand{\mysection}[1]{ \section{#1}}

\title{
Superconducting Proximity Effect on the Edge of Fractional Topological Insulators
}
\author{Meng Cheng}
\affiliation{Condensed Matter Theory Center, Department of Physics, University of Maryland, College Park, MD 20742}
\date{\today}
\begin{abstract}
  We study the superconducting proximity effect on the helical edge states of time-reversal-symmetric fractional topological insulators(FTI). The Cooper pairing of physical electrons results in many-particle condensation of the fractionalized excitations on the edge. We find localized zero-energy modes emerge at interfaces between superconducting regions and magnetically insulating regions, which are responsible for the topological degeneracy of the ground states.  By mapping the low-energy effective Hamiltonian to the quantum chiral Potts model, we determine the operator algebra of the zero modes and show that they exhibit nontrivial braiding properties. We then demonstrate that the Josephson current in the junction between superconductors mediated by the edge states of the FTI exhibit fractional Josephson effect with period as multiples of $4\pi $.
\end{abstract}
\pacs{  05.30.Pr, 03.67.Lx}
\maketitle

\mysection{Introduction}
Topological phases are often characterized by gapless boundary excitations which do not arise in the low-energy part of local lattice Hamiltonians with the same dimension. For example, the integer/fractional quantum Hall states support chiral edge excitations, which have only ``half'' the degrees of freedom as particles moving in a one-dimensional lattice. Recently discovered Topological Insulators(TI) with Time-Reversal(TR) symmetry support gapless helical boundary states in both two and three dimensions~\cite{Hasan_RMP2010, Qi_RMP2011, Kane2005a, Kane2005b,Bernevig2006, Fu_PRL07, Moore_PRB07, Roy_PRB09}. These boundary states can be gapped out by adding certain symmetry-breaking perturbations(e.g. superconducting or magnetic order) which in many cases leads to exotic phases.  A notable example is the $p_x+ip_y$ superconducting state created by superconducting proximity effect on the surface states of three-dimensional TI~\cite{Fu_PRL08}  which exhibits non-Abelian Majorana zero modes in vortices. Similar physics can also be realized on the edges of two-dimensional quantum spin Hall insulators~\cite{Fu_PRB2009}, where Majorana zero modes appear at the interfaces between superconducting and magnetic gapped regions. The Majorana zero modes exhibit unusual properties such as $4\pi$-period Josephson effect~\cite{Kwon_EPJ2003, Fu_PRB2009, Lutchyn_PRL2011} and non-Abelian statistics~\cite{Alicea_NatPhys2011, Clarke_PRB2011, Sau_PRB2011b}, which have important application in quantum information processing~\cite{nayak_RevModPhys'08, Lutchyn_PRL2011, Oreg_PRL2010, Sau_PRA2010}.

On the other hand, strongly correlated topological phases, such as Fractional Quantum Hall(FQH) states~\cite{Tsui_FQHE, Laughlin},  are usually associated with the fractionalization of quantum numbers. It would be even more interesting to study the quantum phases originated from symmetry breaking in the fractionalized boundary states.
In this paper we study superconducting proximity effect on the edge of two-dimensional fractional topological insulators (FTI)~\cite{Bernevig_PRL2006, Karch_PRD2010, Levin_PRL2009, Santos_PRB2011, Neupert_PRB2011},  which can be regarded as the TR symmetric generalization of Laughlin states with filling fraction $\nu=\frac{1}{m}$.   We focus on the properties of the gapless edge states of FTI brought in contact with an s-wave superconductor. 

\begin{figure}[htpb]
  \begin{center}
	\includegraphics[width=0.7\columnwidth]{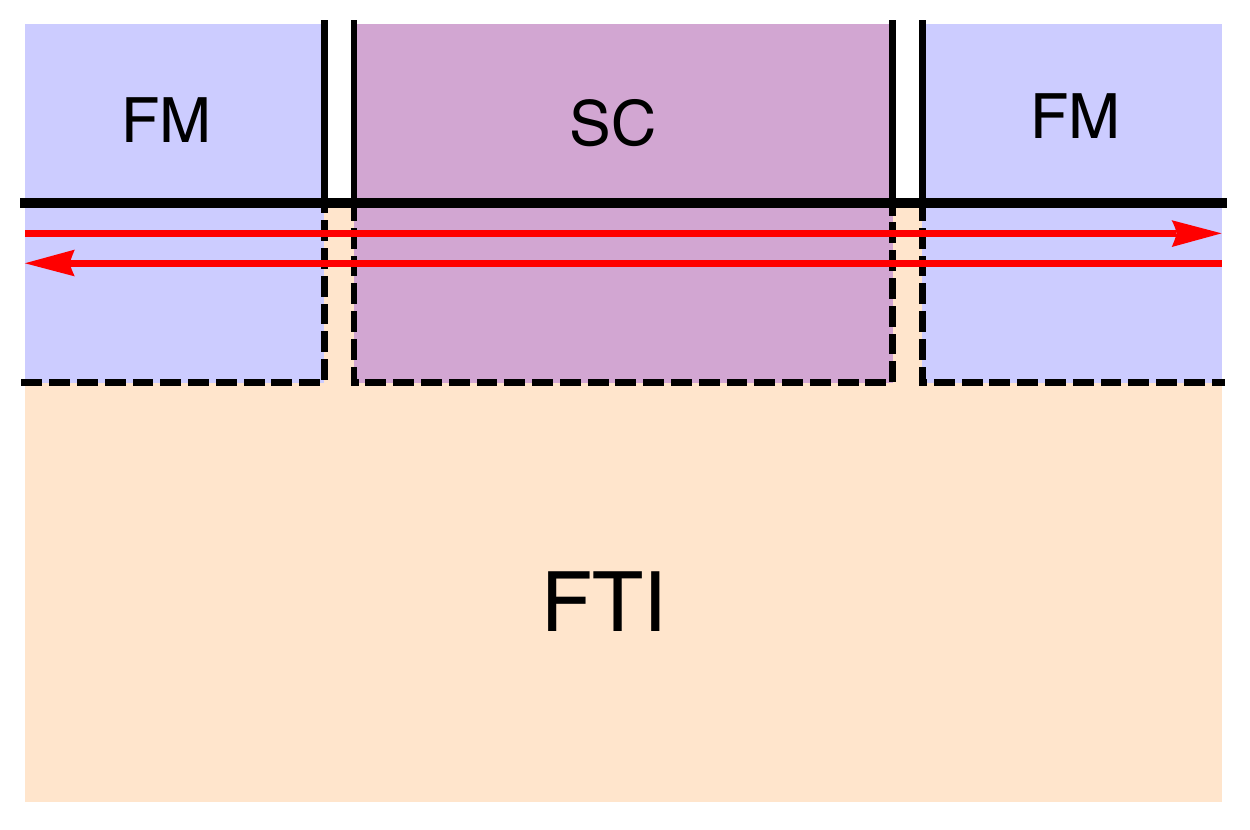}
  \end{center}
  \caption{Schematic illustration of superconductor-ferromagnet junctions on the edge of 2D FTI. Localized zero modes appear at the interface between the superconducting regions(SC) and the magnetically insulating regions(FM).}
  \label{fig:setup}
\end{figure}

Our main finding is that the electron fractionalization drastically changes the superconducting proximity effect on the edge states, as opposed to the non-interacting quantum spin Hall insulators. Most remarkably, domain walls between the superconducting and magnetic gapped regions are found to carry localized zero modes with quantum dimension $d=\sqrt{2m}$. We also determine the operator algebra satisfied by the zero modes which can be regarded as a $\mathbb{Z}_{2m}$ generalization of Majorana fermions. We then argue that the zero modes exhibit non-Abelian statistics upon adiabatic exchanging and determine the braiding matrices. We also discuss possible physical indications of the unusual zero modes and propose an unconventional Josephson effect with period $4\pi m$.

The paper is organized as follows: in Sec. II we review the effective edge theory of $S_z$-conserved FTI. In Sec. III and IV we study the gapped phase driven by s-wave pairing. We reveal the topological degeneracy of the ground states and find the localized zero modes. In Sec. V we discuss the braiding of the zero modes.

\mysection{Edge Theory of FTI}
We start by reviewing the effective edge theory of FTI~\cite{Levin_PRL2009,Santos_PRB2011, Beri_arxiv2011}. We mainly consider FTI with $S_z$ conserved where the two spin species each form Laughlin states with filling fraction $\nu=\frac{1}{m}$ where $m$ is an odd integer, under a spin-dependent magnetic field $\vec{B}=B_0\vec{z}\sigma_z$. It supports gapless edge states described by a helical Luttinger liquid model. Levin and Stern~\cite{Levin_PRL2009} have recently shown that such edge states are protected by TR symmetry if and only if $\sigma_\text{SH}/e^*$ is odd. Here $\sigma_\text{SH}$ is the spin Hall conductance measured in units of $e/2\pi$ and $e^*$ is the elementary charge.  Here the ratio $\sigma_{\text{SH}}/e^*=1$ implying the robustness of the edge states protected by TR symmetry.  The effective Lagrangian density governing the dynamics of the edge states is given by~\cite{Wen_PRB1990, Wen_AdvPhys1995}
\begin{equation}
  \mathcal{L}=\frac{1}{4\pi}\sum_{\sigma=\uparrow,\downarrow} (K_{\sigma\sigma'}\partial_t\phi_\sigma\partial_x\phi_{\sigma'}-V_{\sigma\sigma'}\partial_x\phi_\sigma\partial_x\phi_{\sigma'}).
  \label{}
\end{equation}
The $\vec{K}$ matrix is $\vec{K}=m\sigma_z$ and  $\vec{V}$ is the renormalized charge velocity matrix. The chiral bosonic fields $\phi_\sigma$ satisfy Kac-Moody algebra
\begin{equation}
  [\phi_\sigma(x),\phi_{\sigma'}(x')]=(\sigma_z)_{\sigma\sigma'}\frac{i\pi}{m}\text{sgn}(x-x').
  \label{}
\end{equation}
To simplify our derivation, we define $\varphi=\frac{m}{2}(\phi_{R\uparrow}+\phi_{L\downarrow}), \theta=\frac{1}{2}(\phi_{L\downarrow}-\phi_{R\uparrow})$. They then satisfy the canonical commutation relation
  $[\varphi(x), \partial_{x'}\theta(x')]=i\pi\delta(x-x')$.
The Hamiltonian of the edge theory can be expressed as 
\begin{equation}
  H=\int\di x\,\frac{u}{2\pi}\big[mg(\partial_x\theta)^2+(mg)^{-1}(\partial_x\varphi)^2\big],
  \label{}
\end{equation}
which describes a Luttinger liquid with Luttinger parameter $mg$. Here $g=1$ if $V_{\uparrow\downarrow}=0$, and $g>1(<1)$ if $V_{\uparrow\downarrow}<0(>0)$.

In this bosonic theory, the electron density $\rho_\sigma$ is expressed as $\rho_\sigma=\frac{1}{2\pi}\partial_x\phi_\sigma=\frac{1}{m\pi}\partial_x\varphi$. The physical electron creation operators are given by $\psi^\dag=\frac{1}{\sqrt{2\pi\alpha}} e^{im\sigma_z\phi}$.  Here $\alpha$ is a regularization factor. Notice that proper Klein factors should be included in the expression to ensure the correct fermionic commutation relations but it turns out that they are not relevant for our discussion below, so we omit them to simpify the notations. We also omit the spin indices of the bosonic fields since they are locked to the chiral indices. We adopt the convention that under TR transformation $\phi_\uparrow\rightarrow \phi_\downarrow,\phi_\downarrow\rightarrow \phi_\uparrow-\frac{\pi}{m}$ to guarantee that $\psi_\uparrow\rightarrow \psi_\downarrow,\psi_\downarrow\rightarrow -\psi_\uparrow$. Consequently, $\theta\rightarrow -\theta-\frac{\pi}{2m}$. One can also define $\psi_\text{qp}^\dag\sim e^{i\sigma_z\phi}$ which creates a charge $e/m$ quasiparticle(QP) on the edge.

 The proximity effect of an s-wave superconductor can be taken into account by adding the following pairing term to the edge theory: 
 \begin{equation}
   \mathcal{H}_\text{SC}=\Delta e^{i\chi}\psi_{R\uparrow}^\dag\psi_{L\downarrow}^\dag+\text{h.c.} .
   \label{}
\end{equation}
Here $\Delta$ is the induced superconducting gap and $\chi$ is the superconducting phase. 
Using the bosonic representation, we can write the pairing term as
\begin{equation}
  \mathcal{H}_\text{SC}= \frac{\Delta}{2\pi\alpha} e^{i\chi}e^{im(\phi_L-\phi_R)}+\text{h.c.}=\frac{\Delta}{\pi\alpha}\cos(2m\theta+\chi).
  \label{}
\end{equation}

We first assume $\Delta$ is a weak perturbation  and study its fate using perturbative Renormalization Group(RG) analysis. The flow of the coupling $\Delta$ under RG is given by
\begin{equation}
  \odiff{\Delta(\ell)}{\ell}=\Big( 2-\frac{m}{g} \Big)\Delta(\ell).
  \label{}
\end{equation}
Here $\ell=\ln(l/l_0)$ is the dimensionless flow parameter. The pairing term is relevant when $g>\frac{m}{2}$.  For $m\ge 3$, in the ``non-interacting'' case $g=1$  the pairing is irrelevant. This is a direct consequence of the electron fractionalization. However, we should remark here that although the term is perturbatively irrelevant when $g<\frac{m}{2}$, it can still gap out the edge Luttinger liquid if the bare value of $\Delta$ is large enough when the perturbative RG analysis breaks down.

\mysection{Topological Degeneracy in the Gapped Phase}
We now consider the gapped phase resulting from the mass term $\cos(2m\theta+\chi)$. We choose the gauge $\chi=0$. Semi-classically, $\theta$ is pinned to the minima of $\Delta\cos 2m\theta$ located at:
\begin{equation}
  \theta_n=\frac{\pi}{2m}+\frac{n\pi}{m}, n=0,1,\dots, 2m-1.
  \label{}
\end{equation}
Therefore there are $2m$ degenerate ground states which we denote by $\ket{\theta_n}$. We notice that when $m=1$ it reduces to the well-known topological degeneracy of one-dimensional Majorana chain~\cite{Fidkowski_PRB2011b, Sau_PRB2011, Cheng_PRB2011b}. Since under TR $\theta$ goes to $-\theta$, the semiclassical ground states $\ket{\theta_n}$ are not TR invariant when $m>1$. 

However, it is important to realize that the one-dimensional system under consideration is already the boundary of a 2D system and thus it must be a closed manifold without any boundaries.  If the whole edge is gapped out by the superconducting proximity effect, there is a unique ground state fixed by the boundary condition. The ground state degeneracy only occurs in an open geometry~\cite{Fidkowski_PRB2011b}. To effectively create boundaries on the edge, we need to introduce a different mass term, e.g. TR-breaking perturbations by applying Zeeman field or  proximity effect to a ferromagnetic insulator  to gap out some regions of the edge liquid. See Fig. \ref{fig:setup} for an illustration of the setup. 
To be specific, we consider inducing ferromagnetic order in some region of the edge, by adding the following mass term to the Hamiltonian:
\begin{equation}
  \mathcal{H}_{\text{FM}}= \Delta_\text{FM}\psi_{R}^\dag \psi_L+\text{h.c.} = \frac{\Delta_\text{FM}}{\pi\alpha} \cos 2\varphi.
  \label{}
\end{equation}
Now consider a setup in which the edge is divided into $2n$ segments, $n$ of which are gapped out by the superconducting proximity effect and the others by proximity to ferromagnets, arranged in an alternating order. From our semiclassical analysis, each superconducting segment has $2m$ ground states. Superficially there would be totally $(2m)^{n}$-fold degeneracy. However, due to the global conservation law of QP number mod $2m$ (see below), the degeneracy is actually reduced to $(2m)^{n-1}$. Since the entire edge is gapped away from the interfaces between different regions, the ground state degeneracy can only arise from zero modes localized at the interfaces.  Because there are $2n$ interfaces, each zero mode has a quantum dimension $d=\sqrt{2m}$. 


To understand the nature of the degenerate ground states, we notice that the total charge density on the edge is given by $\rho=\frac{1}{2\pi m}\partial_x \varphi$. Its commutation relation with $\theta$ is $[\rho(x), \theta(x')]=\frac{i\pi}{m}\delta(x-x')$. Therefore the following relation can be derived:
\begin{equation}
  e^{2\pi i Q}\theta e^{-2\pi i Q}=\theta+\frac{\pi}{m}.
  \label{}
\end{equation}
Here $Q=\int\di x\,\rho(x)$ which counts the number of QP's residing on the edge. We can then linearly superpose  $\ket{\theta_j}$ to obtain eigenstates of $e^{2\pi i Q}$:
\begin{equation}
  \ket{n}=\frac{1}{\sqrt{2m}}\sum_{j=0}^{2m-1}\omega^{nj}\ket{\theta_j},\, Q\ket{n}=\omega^n\ket{n}.
  \label{}
\end{equation}
Here $\omega=e^{\frac{i\pi}{m}}$. Therefore, the $2m$ degenerate ground states have different mod $2m$ QP number. This is analogous to the two-fold ground state degeneracy of a one-dimensional class D topological superconductor, distinguished by the global fermion parity.  
Using the TR transformation of $\theta$, one can see that $\ket{n}\rightarrow \ket{2m-n}$ under TR. Therefore except the states $\ket{0}$ and $\ket{m}$, all other states are {\it not} TR invariant.

One may wonder whether the $2m$-fold degeneracy is protected, and if that is the case, what physical property protects such degeneracy.
First of all, the existence of gapless edge modes requires TR symmetry in the bulk of the FTI~\cite{Levin_PRL2009}. Given the TR symmetry,
because the FTI state is fully gapped, at energies well below the bulk gap, QP tunnelings through the bulk are highly suppressed. Due to the fractional statistics of the QP in FTI with statistical angle $\pm\pi/m$, any external physical perturbations must change the QP number of each species by $m$ since they have to be local with respect to electrons. This fact, together with the conservation of the total fermion parity in the gapped superconducting systems, implies an emergent conservation of QP numbers module $2m$ on the edge. Therefore, the degeneracy of the $2m$ ground states with different QP numbers mod $2m$ can not be lifted by any physical perturbations. We therefore conclude that the $2m$-fold degeneracy is protected by the topological order in the bulk FTI and the fermion parity conservation.

\mysection{Low-Energy Effective Hamiltonian and Zero-energy Boundary Modes}
In the previous section we use semiclassical analysis to study the ground state properties of the gapped phase.  We argue that it is natural to relate the ground state degeneracy to localized zero modes on the SC-FM interfaces where the spectra gap has to close. Heuristically, the zero modes are transformations between the ground states.  In this section we put these considerations on a firmer ground by deriving an effective Hamiltonian to describe the low-energy quantum fluctuations around the semiclassical ground states.

First it will turn out to be convenient to regularize the bosonic Hamiltonian on a lattice (after proper rescaling of space):
\begin{equation}
  H=u\sum_i \big[\!-\!g\cos(\theta_i-\theta_{i+1})+\frac{g}{2}n_i^2+\frac{\Delta}{\pi u}\cos 2m\theta_i\big].
  \label{eqn:clock}
\end{equation}
Here $n_i$ is the canonical conjugate operator to the phase variable $\theta_i$: $[\theta_i, n_j]=i\delta_{ij}$. It can be regarded as the discrete version of the QP density. In the limit of large $\Delta$,  $\theta$ can only takes value in a discrete set given by $\theta_n, n=0,1,\dots, 2m-1$. We introduce a set of basis $\ket{\theta_n}$ on each site and the first term $\cos(\theta_i-\theta_{i+1})$ becomes $U_i^\dag U_{i+1}$ where the $U_i$'s have matrix representation 
\begin{equation}
  U_i=\text{diag}(1, \omega, \omega^2, \dots, \omega^{2m-1}),
  \label{}
\end{equation}
which is simply the operator $e^{i\theta_i}$ projected onto the low-energy sector.

The $\sum_in_i^2$ term causes transitions between the $\ket{\theta}$ states. We rewrite  this term  in the $\ket{\theta}$ basis as (on one site)
\begin{equation}
  n^2=\sum_{j=1}^m a_j(V^j+{V^j}^\dag),\: a_j \propto \sum_k k^2 \omega^{kj}.
  \label{}
\end{equation}
Here the operator $V$ is defined as
\begin{equation}
  V_i =
  \begin{pmatrix}
	0 & 1  & 0 & \dots & 0\\
	0 & 0  &  1 & \dots & 0\\
	\vdots & \vdots &\vdots  & \ddots & \vdots \\
	0 & 0 & 0 & \dots & 1 \\
	1 & 0  & 0 & \dots &  0\\
  \end{pmatrix}. 
  \label{}
\end{equation}
We now make the approximation to keep only the $j=1$ term in the sum: $n_i^2\propto V_i+V_i^\dag$. This is justified in the limit of small fluctuations of $\theta$. 
It is worth mentioning that $U$ and $V$ form a $2m$-dimensional representations of the Weyl group algebra:
\begin{equation}
  U^{2m}=1, V^{2m}=1, V U=\omega U V.
  \label{}
\end{equation}
So far we have succeeded in mapping the model \eqref{eqn:clock}  to the $2m$-states quantum Potts model~\cite{Elitzur_PRD1979}:
\begin{equation}
  H_\text{Potts}=-\sum_{i}[(U_i^\dag U_{i+1}+\text{h.c.})+ \lambda(V_i+V_i^\dag)]
  \label{eqn:chiralpotts}
\end{equation}
Here $\lambda\propto g^2$.

We now briefly discuss the symmetry of the Potts model. The Hamiltonian \eqref{eqn:chiralpotts} apparently has a global $\mathbb{Z}_{2m}$ symmetry given by
\begin{equation}
  Q=\prod_{i}V_i^\dag.
  \label{}
\end{equation}
To see the physical meaning of the $\mathbb{Z}_{2m}$ symmetry, it is useful to go to a ``dual'' basis $\ket{n}$ on each site which are the eigenstates of $n$. In this basis, the global symmetry takes a very transparent form: $
Q=\omega^{\sum_i n_i}$
which is nothing but the mod $2m$ QP number we have mentioned in the previous section.


A crucial property of the Potts model is the self duality~\cite{Savit_RMP1980}, revealed by the duality transformation~\cite{duality_advphys2011}:
\begin{equation}
  U_{i-1}^\dag U_i=\tilde{V}_i^\dag, \tilde{V}_i=U_i^\dag U_{i+1},
\end{equation}
  which maps $H_\text{Potts}$ into itself with $g\rightarrow g^{-1}$ up to an overal rescaling.
We can use the results in [\onlinecite{duality_advphys2011, Rajabpour_JPA2007, fendley_unpublished}] to obtain the explicit form of the zero modes at the special solvable point $g=0$.  Notice the product of the original variable and the neighbouring dual variable $\gamma_j=U_i \tilde{U}_{i-1}^\dag$ satisfies the following algebra:
\begin{equation}
  \gamma_i\gamma_j=\omega\gamma_j\gamma_i,\: i< j, \gamma_i^{2m}=\pm 1,\gamma_i^\dag\gamma_i=1.
  \label{eqn:para}
\end{equation}
For $m=1$ this is just the algebra of Majorana fermions. We call the zero modes $\gamma$'s which satisfy \eqref{eqn:para} as $\mathbb{Z}_{2m}$ zero modes. Using this result, the original model can be rewritten in terms of the $\gamma_i$ operators~\cite{fendley_unpublished}:
\begin{equation}
  H_\text{Potts}=e^{\frac{\pi i}{2m}}\gamma_{1,i+1}^\dag\gamma_{2,i}+\lambda e^{\frac{\pi i}{2m}}\gamma_{1,i}^\dag\gamma_{2,i}+\text{h.c}.
  \label{}
\end{equation}
The explicit expression of the $\gamma$'s in terms of the original variables are given below:
\begin{equation}
  \gamma_{1,i}=U_i\prod_{j<i}V_j^\dag, \gamma_{2,i}=e^{-\frac{\pi i}{2m}}U_i\prod_{j\leq i}V_j^\dag,
  \label{}
\end{equation}
which generalizes the Jordan-Wigner transformation for fermions. Notice that we have multiplied a phase factor $e^{\frac{\pi i}{2m}}$ to make sure that $\gamma_{i}^{2m}=1$.

For a finite chain with sites starting at $i=0$ and ending at $i=N$, $\gamma_L\equiv\gamma_{1,i=0}$ and $\gamma_R\equiv\gamma_{2,i=N}$ are decoupled from the Hamiltonian at $\lambda=0$, similar to Kitaev's Majorana chain model~\cite{Kitaev_Majorana}. One can also check that $\gamma_{L, R}$ commute with all other terms in the Hamiltonian.  In terms of the original bosons, the two zero modes are expressed as
\begin{equation}
\gamma_L\sim e^{i\theta_{i=0}}, \gamma_R\sim e^{i\theta_{i=N}}Q.
\label{eqn:zeromode}
\end{equation}
When $\lambda\neq 0$, the model in general can not be solved exactly . However, since when $\lambda=0$ the bulk is fully gapped, one can imagine slowly turning on a small but finite $\lambda$ without closing the bulk gap. The finite $\lambda$ phase should be adiabatically connected to the gapped phase of the FTI edge theory. During this adiabatic evolution, the zero modes are renormalized but remains localized as long as the bulk gap is not closed. In particular, we expect the relation \eqref{eqn:zeromode} holds true at least in the low-energy sector.  
\comments{
It is useful to work out the action of the zero modes on the ground states with fixed quasiparticle numbers:
\begin{equation}
  \begin{gathered}
  \gamma_{L}\ket{n}=\ket{n+1}\\
  \gamma_{R}\ket{n}=e^{\frac{ i(n+1)\pi}{m}}\ket{n+1}.
\end{gathered}
  \label{}
\end{equation}
}
We also identify from \eqref{eqn:zeromode}
\begin{equation}
  Q=e^{\frac{i\pi}{2m}}\gamma_{L}^\dag\gamma_{R}.
  \label{}
\end{equation}
Again  we expect it to be valid universally in the gapped phase.


We derive the expression of the zero modes $\gamma_i$ within the low-energy effective Hamiltonian which takes a non-local form with a string operator attached. One may wonder whether they are local objects in terms of the QP operators of the FTI edge theory. Let us remark that the local fields $U_i$ and $V_i$ in the effective Hamiltonian are themselves non-local in the original edge theory. Although a formal proof is lacking at this stage,  we believe it is highly likely the zero modes are {\it local } in terms of the quasiparticle operators.

\mysection{Braiding of the $\mathbb{Z}_{2m}$ Zero Modes}
We now discuss the braiding statistics of the $\mathbb{Z}_{2m}$ zero modes. We need to specify what braiding means for the zero modes localized at the domain walls in one dimension. In a closely related context, it has been recently demonstrated~\cite{Alicea_NatPhys2011, Clarke_PRB2011, Sau_PRB2011b} that Majorana zero modes in one-dimensional wires can be adiabatically exchanged either by forming networks out of the wires and moving the Majorana zero modes using the ``T''-junction geometry, or by a series of QP tunneling processes. The non-Abelian braiding statistics resulting from these operations takes the same form as the one of the Majorana fermions in two-dimensional $p_x+ip_y$ superconductors up to an undetermined overall phase. This paves the way to perform topological quantum computing in quantum wires.
There is no fundamental difficuties in exploiting these ideas to the situations at hand and therefore we can discuss the braiding of $\mathbb{Z}_{2m}$ zero modes.

In the following discussion, we derive possible forms of the braiding matrix based on general principles without appealing to details of the implementation of braidings. 
We will extensively use the language of the algebraic theory of anyons (namely, modular tensor category theory) and assume that the braiding matrices of the zero modes satisfy some of the basic relations originally developed for anyons in topological phases in two dimensions~\cite{Kitaev2006}. Whether this description applies to quasi-one-dimensional objects as those considered here is not justified {\it a priori}. It is an important question to clarify the relationship between the anyon model at hand and the modular tensor category theory describing two-dimensional topological phases.  The information we  need is the fusion algebra of the anyons as well as the monodromy equation relating the braiding matrices to the topological spins of the particles.

We start from the fusion algebra of the $\mathbb{Z}_{2m}$ zero modes. There are $2m+1$ types of particles, denoted by $\sigma, \psi_k$ where $k=0,1,\dots, 2m-1$. Here $\psi_0\equiv 1$ is the vacuum. The nontrivial part of the fusion algebra is given by
\begin{equation}
  \begin{gathered}
  \sigma\times\sigma=\sum_{k=0}^{2m}\psi_k,\, \sigma\times\psi_k=\psi_k\\
\psi_p\times\psi_q=\psi_{p+q\text{ mod } 2m}.
\end{gathered}
  \label{eqn:fusion}
\end{equation}
Here $\sigma$ represents the $\mathbb{Z}_{2m}$ zero mode and $\psi_n$ is the state with $n$ QPs, which are Abelian with quantum dimension $1$. The vaccum fuses trivially with everything else. The fusion of two $\sigma$'s follows straightforwardly from the topological ground state degeneracy discussed in Sec. III. The fusion of $\psi$ particles follows from the fact that they are labeled by the number of QP's mod $2m$ in the superconducting segment of the FTI edge, and thus naturally form a $\mathbb{Z}_{2m}$ structure.

Using the language of modular tensor category, we have the following basis data from the fusion algebra:
\begin{equation}
 N^{\psi_n}_{\sigma\sigma}\equiv\text{dim}\,V^{\psi_n}_{\sigma\sigma}=1.
  \label{eqn:fusiondim}
\end{equation}
We have already counted the quantum dimensions of the anyons: $d_{\sigma}=\sqrt{2m}, d_{\psi_n}=1$.

Now let us consider braiding a pair of $\sigma$ particles labeled as $1$ and $2$ and for simplicity we assume that physically they belong to the same topological segment. As discussed in the previous analysis, such a segment has in fact only one ground state within a given mod $2m$ QP number sector. This translates into the one-dimensionality of the fusion space as given in \eqref{eqn:fusiondim}. Therefore the result of an adiabatic exchange is essentially Abelian: the ground states in different mod $2m$ QP number sector can not mix and each acquires an Abelian Berry phase. Let us assume that the ground state with mod $2m$  QP number $n$ acquires a phase $\alpha_n$. In  more abstract terms, the $R$-matrix encoding the effects of braiding $R^{\sigma\sigma}_{\psi_n}$ is actually one-dimensional since $N^{\psi_n}_{\sigma\sigma}=1$. Choosing a proper normalization, it is exactly the phase factor $e^{i\alpha_n}$. 

Without knowing how the braidings are implemented, one can actually determine the $R$-matrix to a large extent using the monodromy equation in our case when fusion spaces are all one-dimensional. The operation of two consecutive braidings, being equivalent to moving one anyon around another, is termed as a monodromy. It is known that the monodromy is fully characterized by the topological spins of the anyons. Using the fusion algebra \eqref{eqn:fusion}, we have the following relation between the $R$-matrix and the topological spins of the anyons $\sigma$ and $\psi_n$~\cite{Kitaev2006}:
\begin{equation}
  (R^{\sigma\sigma}_{\psi_n})^2=\frac{\theta_{\psi_n}}{\theta_\sigma^2}.
  \label{}
\end{equation}
To figure out the topological spin of $\psi_n$, we notice that $\psi_n$ can be regarded as a composite of $n$ QP. Each QP, being an Abelian anyon in a $\nu=1/m$ Laughlin FQH liquid, has topological spin $e^{\frac{is\pi}{m}}$ where $s\in \mathbb{Z}$. Thus the topological spin $\theta_{\psi_n}=e^{\frac{isn^2\pi}{m}}$.~\cite{preskill_notes}  Since $(R^{\sigma\sigma}_{\psi_n})^2=e^{2i\alpha_n}$, we find
\begin{equation}
  e^{i\alpha_n}=\pm \theta_\sigma^{-1}e^{\frac{i\pi n^2 s}{2m}}.
  \label{eqn:braiding1}
\end{equation}
Therefore, we have determined $e^{i\alpha_n}$ up to a  global phase $\pm\theta_{\sigma}^{-1}$. This result is derived from very general consideration and should be independent of the implementations of the braiding operations. One can see that if $m=1$ and $s=1$, \eqref{eqn:braiding1} gives the well-known Berry phases for the even and odd parity ground states of two Majorana zero modes~\cite{Ivanov_PRL'01} if we choose all the unspecified sign factors in \eqref{eqn:braiding1} to be $1$.

We go on to consider four zero modes $\gamma_{1,2,3,4}$ arranged such that $\gamma_i\gamma_j=e^{\frac{i\pi}{m}}\gamma_j\gamma_i$ for $i<j$. We assume that $\gamma_1$ and $\gamma_2$ belong to the same topological segment and the same to $\gamma_3$ and $\gamma_4$. Naturally we choose the Fock basis formed by the degenerate ground states of the two segments, labeled by mod $2m$ QP number $Q_1\equiv Q_{12}\propto\gamma_1^\dag \gamma_2, Q_2\equiv Q_{34}\propto\gamma_3^\dag\gamma_4$: $\ket{n_1, n_2}=\gamma_1^{n_1} \gamma_3^{n_2}\ket{0},n_1, n_2=0,1,\dots, 2m-1$. Braiding $\gamma_1$ and $\gamma_2$ (or $\gamma_3$ and $\gamma_4$) just generates a Abelian phase on the Fock states. The nontrivial one is braiding $\gamma_2$ and $\gamma_3$ which belong to different topological segments. To derive the non-Abelian Berry phase acting on the ground state manifold, we make use of the $\mathbb{Z}_{2m}$ conservation of QP number. After braiding, the QP number $Q_1'\propto\gamma_2\gamma_3^\dag$ as well as $Q_2'\propto\gamma_1^\dag\gamma_4$ should remain the same. This implies that if one rotates the Fock basis to one in which $Q_{1,2}'$ becomes diagonal,  the braiding results in an Abelian Berry phase as discussed above. Then one can unwind the basis transformation to obtain the braiding matrix in the original basis.
Similar procedure can be carried out for any $2n$ zero modes, although our method for determining the braid matrix quickly becomes very inefficient.

So far our discussion of braiding has been general. A realistic scheme could possibly be developed exploiting the idea of measurement-only topological quantum computation~\cite{measurement-only}, in which braidings are carried out by a series of measurements of the fusion outcomes for pairs of anyons in the presence of additional topological qubits that are properly initialized. 

\mysection{Fractional Josephson Effect}
In this section we discuss the manifestation of the topological degeneracy in Josephson transport. We show that the Josephson current has a period of $4\pi m$, thus generalizing the $4\pi$ Josephson effect in topological superconductor~\cite{Kwon_EPJ2003, Fu_PRB2009}. 

Before we proceed, it is useful to understand the gauge transformation on the state $\ket{n}$. Assuming that we change the phase $\chi$ by $2\pi$: $\chi\rightarrow \chi+2\pi$, then $\theta_j\rightarrow \theta_j+\frac{\pi}{m}=\theta_{j+1}$. Therefore
$\ket{n}\rightarrow \omega^{-n}\ket{n}$, implying $\gamma\rightarrow \omega\gamma$.

In terms of the localized modes, we can write down the general form of the effective Hamiltonian in the tunneling regime to leading order in the tunneling amplitude $t$:
\begin{equation}
  H_\text{eff}=\Gamma(\chi)\gamma_{L}^\dag \gamma_{R}+\text{h.c.}.
  \label{}
\end{equation}
Here $\gamma_{L}$ and $\gamma_{R}$ are localized zero modes at the two ends of the junction. Notice that here it is crucially important that the Josephson current is carried by quasiparticles instead of electrons. Although the junction is gapped out by TR-breaking field (e.g. ferromagnet), it is still part of the fractional quantum spin Hall fluid. Since they carry charge $e/m$, the period of the Josephson current should become $4\pi m$. In the following we will provide a more rigorous calculation of the $4\pi m$ period of the Josephson current.

The DC Josephson current is then given by
\begin{equation}
  I(\chi)=2e\frac{\di \langle H_\text{eff}\rangle}{\di \chi}=4e\,\Re\big[\odiff{\Gamma}{\chi} \langle\gamma_L^\dag \gamma_R\rangle\big].
  \label{}
\end{equation}
Here $\langle\gamma_L^\dag \gamma_R\rangle$ is the conserved mod $2m$ QP number. Therefore the periodicity of $I$ is completely determined by the periodicity of the phase-dependent coupling $\Gamma(\chi)$.

Now we consider the coupling $\Gamma(\chi)$. We assume a gauge choice in which superconducting phase to the left of the junction is fixed and the the phase on the right is $\chi$. Now we increase the superconducting phase by $2\pi$: $\chi\rightarrow \chi+2\pi$. As we just demonstrate, the zero modes $\gamma_L\rightarrow \gamma_L, \gamma_R\rightarrow \omega\gamma_R$. Equivalently, this extra phase factor can be absorbed into the coupling $\Gamma$:
\begin{equation}
  \Gamma(\chi+2\pi)=\omega \Gamma(\chi).
  \label{}
\end{equation}
Consequently, the Josephson current
\begin{equation}
  I(\chi+2\pi) = 4e\Re\Big[\omega \Gamma'(\chi)\langle\gamma_L^\dag \gamma_R\rangle\Big],
  \label{}
\end{equation}
which is in general different from $I(\chi)$. 
 It is easy to see that for $I(\chi)$ to return to its initial value, $\chi$ has to advance by $4\pi m$ since $\omega^{2m}=1$. So in summary we have derived that
\begin{equation}
  I(\chi+2\pi)\neq I(\chi),\: I(\chi+4\pi m)=I(\chi).
  \label{}
\end{equation}
Although we have not determined the precise functional dependence of $I(\chi)$, it is sufficient to conclude that the Josephson current has a period $4\pi m$.


We now briefly discuss the AC Josephson effect. The conventional AC Josephson current at finite voltage $V$ has frequency $\frac{2eV}{\hbar}$. In our case, since the Josephson current is transported by a single fractionalized QP, the AC Josephson current has frequency $\frac{eV}{m\hbar}$. When the Josephson junction is irradiated by microwaves with frequency $\omega$, the Shapiro steps in DC current are observed at $V_n=\frac{nm\hbar\omega}{e}$, i.e. we only see Shapiro steps at multiples of $2m$. However, the missing steps are filled in by higher-order tunneling terms in the junction. A recent proposal of a three-leg Shapiro-step measurement may also be adopted here to overcome this obstacle~\cite{Jiang_arxiv2011}.

\mysection{Conclusion and Discussion}
In conclusion, we have investigated the exotic gapped phase formed on the edge of FTI under superconducting proximity effect. This phase is characterized by topological degeneracy determined by the electron fractionalization. The degenerate ground states are distinguished by the mod $2m$ QP number. We relate the degeneracy to $\mathbb{Z}_{2m}$ zero modes on the boundary of the fractionalized topological superconducting region. We derive an effective Hamiltonian to describe low-energy physics which maps to a quantum Potts model and allows for explicit construction of the zero modes. We then discuss the braiding of these zero modes. We also propose fractional Josephson effect as a signature of this unusual phase.

It will be very interesting to work out the complete algebraic description of the non-Abelian excitations found in this work, in the framework of tensor category theory. One may wonder, based on the relation to the chiral Potts model, that the zero modes discussed in this work can be regarded as one-dimensional analog of $\mathbb{Z}_{2m}$ parafermions
arising in, e.g. Read-Rezayi fractional quantum Hall states~\cite{Read_PRB99}. However, as non-Abelian anyons, they have apparently different fusion rules and braiding statistics and therefore should not be confused with each other. It is indeed true that both of them have close relation to $\mathbb{Z}_{2m}$ clock models. The wavefunctions of parafermionic fractional quantum Hall states are constructed from the correlation functions in parafermion conformal field theory~\cite{FZ} which describes the critical point of certain $\mathbb{Z}_{2m}$-symmetric statistical model. The zero modes studied in this work are related to the ordered phase of chiral Potts model. In both cases, the zero modes and parafermions are expressed by the order and disorder variables of a $\mathbb{Z}_{2m}$-symmetric model~\cite{Rajabpour_JPA2007}. However, they should be placed properly in very different phases of the model.

  We also notice that non-Abelian anyons with quantum dimension $d=\sqrt{N}$ where $N>1, N\in \mathbb{Z}$ have been found as lattice dislocations in $\mathbb{Z}_N$ gauge theory~\cite{You_arxiv2012} and also lattice fractional quantum Hall states~\cite{Berkeshii_arxiv2011}. The topological field theory of non-Abelian phases containing these anyons has not been found. 
  Some candidates include the $\mathbb{SO}(N)_2$ Chern-Simons theory~\cite{Hastings_arxiv2012}  and a
  $\mathbb{U}(1)\times\mathbb{U}(1)\rtimes \mathbb{Z}_2$ Chern-Simons theory studied by Barkeshli and Wen~\cite{Barkeshli_arxiv2010} may be relevant in this context.

We now discuss possible future directions. We have considered the edge states of 2D FTI. A natural question is whether the approach taken in this work can be generalized to fractional topological insulators in three dimensions~\cite{Maciejko_PRL2010, Swingle_PRB2011} the surface states of which are fractionalized, helical non-Fermi liquid. In contrast to the 2D case where at least theoretically the existence of FTI has been firmly established, in 3D the situation becomes much more complicated and so far apart from one exactly solvable model~\cite{Levin_PRB2011}, other attempts are all based on parton constructions and gauge theories. However, one could expect that superconducting proximity effect on such exotic surface states would lead to very interesting phenomena. In particular, one can suspect that there may be unusual zero modes localized in the cores of superconducting $hc/2e$ vortices. 

{\it Note added.} Upon finishing the manuscript we become aware of several preprints ~\cite{Clarke_arxiv2012, Linder_arxiv2012, Vaezi_arxiv2012} on closely related topics. 

\mysection{Acknowledgement}
We thank Kai Sun for stimulating discussions at the initial stage of this work. We are grateful to Zheng-Cheng Gu, Hong-Hao Tu, Zhenghan Wang, Chetan Nayak and Roman Lutchyn for enlightening conversations. We thank Xin Wang for valuable comments on the manuscript. We also thank Microsoft Station Q for hospitality during the finalization of the manuscript.

\end{document}